\documentclass[prl,aps,twocolumn,showpacs]{revtex4}
\input epsf
\usepackage{graphics}
\usepackage{graphicx}

\begin{document}
\title{Measurement of an integral of a classical field with a single quantum particle.}

\author{ Lev Vaidman and Amir Kalev}


\affiliation
{ School of Physics and Astronomy\\
Raymond and Beverly Sackler Faculty of Exact Sciences\\
Tel-Aviv University, Tel-Aviv 69978, Israel}

\date{}

\vspace{.4cm}

\begin{abstract}
  A method for measuring an integral of a classical field via local
  interaction of a single quantum particle in a superposition of
  $2^N$ states is presented. The method is as efficient as a quantum method
  with $N$ qubits passing through the field one at a time and it is
  exponentially better than any known classical method  that
  uses $N$ bits passing through the field one at a time.
  A related method for searching a string with a quantum particle is proposed.
\end{abstract}


\maketitle For the past two decades, we are witnessing dramatic growth
of research in the field of quantum information: analysis of
information tasks that can be performed more efficiently using quantum
devices, for example, fast computation \cite{DJ,Shor} and fast
searching \cite{Gr96,Gr97}.  Since quantum ``hardware'' used for
storing, transmitting, and manipulating information is usually very
different from its classical counterpart, there is no unique  way to
make the comparison between quantum and classical systems.  It has become
customary to measure quantum and classical information systems by
comparing the number of basic data storage units - namely, qubits and
bits respectively- needed for a particular task.  However, other
aspects may prove to be also important.  For example, due to
difficulties in arranging direct photon-photon interactions,
 an extensive research of what can be achieved using linear
optical devices was done \cite{KLM}.  Thus, the number of qubits stored in the
Hilbert space of the quantum system performing the information task,
is not always the only (or the best) measure by which we can evaluate
the efficiency of a quantum system. Depending on the possibility of
practical applications, various quantum schemes might have particular
advantages.

 Grover's fast search algorithm \cite{Gr96,Gr97} which
uses $N$ qubits can be performed with a  single particle with $2^N$
states \cite{Ld}. Meyer \cite{Mr}, suggested that it can be done for
other tasks too and in this paper we present such modification for a
recently proposed task of measuring an integral of a classical field
using quantum devices.

Recently, a quantum method,  using a single qubit for measuring the parity
of an integral of a classical field, 
\begin{equation}
\label{eq:int}
I=\int_A^B  \phi (x)  dx,
\end{equation}
provided it takes on only positive integral values, has been suggested
\cite{GH}.  This method was generalized, by Vaidman and Mitrani (VM)
\cite{VM}, to compute the value of the integral itself, using $N$
qubits represented by $N$ spin-$1\over 2$ particles (or any other
two-level quantum systems) which are sent one at a time through the
field.  Furthermore, the VM method is applicable when the integral may
take on non-integer values.  The
precision of this method turns out to be exponentially better than any
known classical method which uses $N$ bits sent one at a time.

We will describe how a single (spin-zero) particle, which passes only once through the
field, can be used to evaluate the integral of the field with the same precision as
the VM method. We let the particle  be in $K=2^N$ distinct sites, so  it requires exponentially increasing precision \cite{EJ,TS}.

We outline the algorithm in what follows.
The particle is initially prepared to be in a superposition of equal amplitudes and vanishing relative phases,
over the $K=2^N$ consecutive separate sites:
\begin{equation}
\label{psiin}
| \Psi_{\rm in}\rangle = {1\over {\sqrt K}}\sum_{k=1}^K |k\rangle .
\end{equation}
Next, we send this ``train of amplitudes'' through the field with constant velocity,
see Fig.1.
\begin{figure}[b]
\label{fig1}
\includegraphics[totalheight=1.8cm,width=6.5cm]{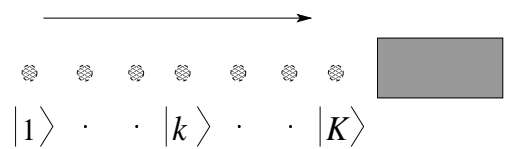}
\caption{The ``particle train'' passes through the field}
\end{figure}

We arrange a local field-particle interaction of the form
\begin{equation}
\label{Hint}
H_{int} =g(x, t) \phi (x) ,
\end{equation}
 in such  a way that the strength of the coupling of  the field to the $k$'th part of
the particle  is  proportional to the index number $k$:
\begin{equation}
\label{couple}
g (x_k(t), t) =\frac{k}{K\alpha},
\end{equation}
where $x_k(t)$ is the location of  $k$'th part at time $t$ and $\alpha$ is
the parameter which we fix depending on the given information
about possible values of the integral of the field.
After the particle completes its passage through the field, its final state (due to the interaction) is
\begin{equation}
\label{psifin}
| \Psi_{\rm fin} (I)\rangle = {1\over {\sqrt K}}\sum_{k=1}^K e^{-i
  {{2\pi kI}\over  K\alpha}} |k\rangle .
\end{equation}
For the special case of $I= \alpha m$, where $m =0, 1, ... K-1$, we
obtain $K$ mutually orthogonal states.  These $K$ states represent a
basis of the Hilbert space of the particle.  Thus, a measurement in
this basis yields the correct value of $m$ with probability 1, exactly
like the VM method does with $N=\log_2 K$ particles.

The VM method \cite{VM} also provides an answer with a good precision
for a more general case when $I$ is not necessarily a multiple of
$\alpha$.  In this case,  the measurement  always yields one
of the discrete values, $\tilde I = \alpha m$, and the probability for
the error, $\delta I = \tilde I -I$, is
\begin{equation}
\label{eq:probdelta}
p(\delta I)=\prod_{n=1}^N \cos^2{{\delta I \pi}\over {2^n \alpha}}.
\end{equation}

In our algorithm we also get one of the values $\tilde I = \alpha m$,
and the probability for the error is given by the squared norm of the
scalar product of the states corresponding to $I$ and  $\tilde I$:
\begin{equation}
\label{eq:probdeltaVK}
 p(\delta
I)= |\langle \Psi (\tilde I)| \Psi (I)\rangle|^2  = \frac{\sin^2{\delta I \pi \over\alpha}}{4^N \sin^2{\delta I \pi
      \over{2^N \alpha}}} .
\end{equation}

Although expressions (\ref{eq:probdelta}) and (\ref{eq:probdeltaVK})
look different, they are, in fact, identical. This can be checked in a
straightforward manner by mathematical induction on $N$.  The equality
is not a coincidence. In fact, from mathematical point of view, the
two methods are isomorphic. We can make the correspondence between the
state $|k\rangle$ and a state of $N$ spin-$1\over 2$ particles which
``writes'' the number $k$ in a binary form with $|{\uparrow}\rangle
\equiv 0$ and $|{\downarrow}\rangle \equiv 1$. We arrange the
interaction between the spins and the field such that the spin
corresponding to $j$th digit accumulate the phase $-{{2\pi 2^j I}\over
  {K\alpha}}$ when the spin is ``down'' and zero phase when the spin
is ``up''. In this way the overall phase of $N$ particles in a state
corresponding to state $|k\rangle$ will be exactly as in our case:
$-{{2\pi kI}\over K\alpha}$. Thus, if we start with $N$ spins
originally pointing in the $x$ direction, i.e. in the state ${1\over
  \sqrt 2}(|{\uparrow}\rangle +|{\downarrow}\rangle )$, then we obtain
the state (\ref{psifin}) after the interaction with the only change
that $|k\rangle$ represents a corresponding state of $N$ spin-$1\over
2$ particles . The interaction which leads to the phase $-{{2\pi 2^j
    I}\over {K\alpha}}$ when the spin $j$ is ``down'' and no phase if
the spin is ``up'', is exactly the magnetic field in the $z$ direction
of the VM method. Therefore, mathematically, the two methods are
equivalent. The implementation, is of course different. It depends on
the physical system, what is  easier: sending $N$ spins one at a
time or sending the train of $2^N$ wave packets.

The function $ p(\delta I)$ is exactly the interference pattern of
$K=2^N$ slits, see Fig. 2. It becomes well localized with large $K$,
but it is periodic with period $\alpha K$. In fact, what is measured
is $I {\rm mod}(\alpha K)$ and the error should be understood as
$(\tilde I -I){\rm mod}(\alpha K)$. Following VM, we consider the
situation in which $I$ is of the order of $M= {{\alpha K}\over {10}}$,
so we can neglect the complications following from the
periodicity of the function $ p(\delta I)$.
\begin{figure}[t]
\includegraphics[angle=-90,totalheight=8cm,width=8cm]{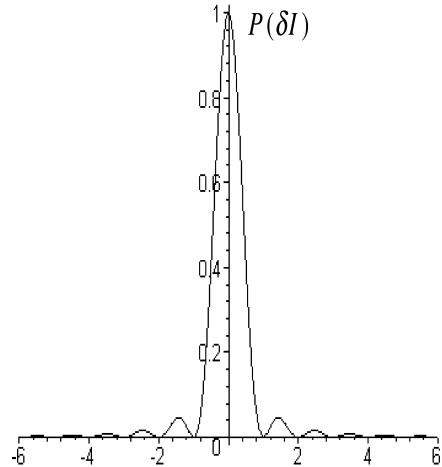}
\caption{The probability of the error $p(\delta I)$ for the quantum methods for $N=7$. (In the figure, $\delta I$ is in units of $\alpha$.)
The outcome of the measurement $\tilde I$ may be one of the values $\alpha m$, so in a particular experiment
$\delta I$ may obtain only a discrete value, $\alpha n + I {\rm mod} \alpha$.}
\end{figure}

The uncertainty of the measurement can be characterized as the standard deviation:
\begin{equation}
\label{DELTA2}
\Delta I = \sqrt {\langle I^2 \rangle - \langle I \rangle^2} \simeq
\frac{10M}{\sqrt{2} \pi}\frac{1}{\sqrt{2^N}}.
\end{equation}
It is also useful to compute another measure of uncertainty, namely,
 the mean  absolute deviation of the measured value
\begin{equation}
\label{DELTA1}
\Delta '  I = \langle|\delta I|\rangle \simeq \frac{10M}{2\pi^2}\frac{\ln 2^N}{2^N}.
\end{equation}

The uncertainty of the corresponding classical method, described in
\cite{VM}, in which $N$ bits are sent one at a time through the field,
is of the order of $1\over \sqrt N$, i.e. it is exponentially larger
than the uncertainty in quantum methods. If we remove the constrain of
sending bits one after the other, we can construct a much better
classical method, but still there is some advantage for the quantum
methods.  In this case the $N$ bits are sent together and they
function as a counter which can go up to $2^N$. If we arrange that the
counter ``clicks", while moving through the field, with probability
\begin{equation}
\label{dp}
dp= \alpha \phi (x) dx,
\end{equation}
 then the resulting  standard deviation $\Delta  I_{cl}  \simeq \sqrt{\frac{10MI}{2^N}} $
 is of the same order as the standard deviation in  quantum methods (\ref{DELTA2}).
However,  the average of the absolute value of the error
\begin{equation}
\label{DELTA1cl}
\Delta ' I_{cl} = \langle|\delta I_{cl}|\rangle  \simeq \sqrt{\frac{2}{\pi}}\Delta I_{cl} =
\sqrt{\frac{20MI}{\pi 2^N}},
\end{equation}
turns out to be larger than  that of the quantum methods (\ref{DELTA1}).

It is interesting to note that a classical algorithm can achieve the
same precision by sending the bits one by one, when local memory is
allowed. We first start with a particle (a ``marker") which goes
through the field and occasionally leaves marks with the same
probability law (\ref{dp}) as our $N$-bit counter. Then, we use our
bits to count the marks.  The counting of the marks can be done in the
following way.  At the beginning, all bits are initialize to 0. The
bits go one after the other along the path. They all behave according
to the following rule: when a bit in a state 0 meets a mark, it erases
the mark and flips to 1, while when a bit in a state 1 meets a mark,
it leaves the mark undisturbed and flips to 0. It is easy to see that
the final states of the bits after they all pass through the marked
path is the binary representation of the total number of marks created
by the marker.

In our method all parts of the wave function of the particle pass
through all points of the field. Can we get some information when
different parts of the wave of the particle pass only through parts of
the field? We cannot find the integral of the field in this way. But
there is a specially tailored task of a similar type which can be
accomplished with a single quantum particle. The classical solution of this task  requires a large number of bits.

\begin{figure}[b]
\label{fig3}
\includegraphics[totalheight=2.0cm,width=5.5cm]{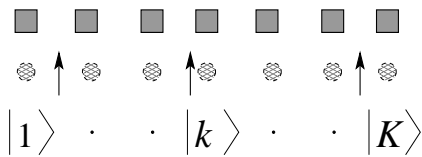}
\caption{A single quantum particle reads a string of $K$ bits.
Each part of the superposition of the particle passes through location of
one of the bits.}
\end{figure}

Consider $K=2^N$ local classical bits which we
want to read.  There are $2^K$ possible strings $\{a_k\}$, but in our
special task we consider a situation in which it is known that our set
of bits can be in one of $N +1$ specific strings. Whatever the strings
are,  in order to find the string, the number of bits we have to approach is larger than
$\log_2 N$ because these bits have to specify the chosen string. Since
in this scenario each particle (and in the quantum analog each part of the
particle wave) approaches only one bit, classically, we need at least $\log_2 N$ particles.
We will show that for a specific set of strings we need just one quantum particle to achieve this goal.

For
$K=16$ our set of  strings is:
\begin{eqnarray}
\label{stringspecial}
\nonumber 0~0~0~0~0~0~0~0~0~0~0~0~0~0~0~0\\
\nonumber 0~0~0~0~0~0~0~0~1~1~1~1~1~1~1~1\\
0~0~0~0~1~1~1~1~0~0~0~0~1~1~1~1\\
\nonumber 0~0~1~1~0~0~1~1~0~0~1~1~0~0~1~1\\
\nonumber 0~1~0~1~0~1~0~1~0~1~0~1~0~1~0~1
\end{eqnarray}


The general rule is clear from the example.
In the $n$th string,
the set of $K\over 2^{n-1}$ bits  0 is followed by  the same number
bits 1, which followed again by the same number of bits  0, etc., until the string ends.

In our quantum method, we again use a single quantum particle prepared
in a superposition of $K$ states without relative phase (\ref{psiin}).
Each part of the superposition passes through location of one of the
bits, Fig 3. The interaction is such that it acquires phase $\pi$ if
the bit is 1 and 0 if the bit is 0. It is easy to see that for
different strings from our special set we obtain in this way mutually
orthogonal states. Thus, we have shown that a single quantum particle
can read reliably $2^N$-bit string provided it is one out of
particular $N +1$ strings.
Using classical devices, for this task we need more than  $\log_2 N$ bits.

It seems that technology today is not at the stage of building a
quantum device which works better than its classical
counterpart. However, experiments, similar to those which show proof
of principle for operating  a quantum computer are certainly capable to
show the proof of principle of the results presented here.

It is a pleasure to thank the members of the Quantum Group at Tel-Aviv
University
for helpful discussions.
This research was supported in part
by grant 62/01 of the Israel Science Foundation.



\end{document}